\documentclass[preprint,prc,aps]{revtex4}
\everymath={\displaystyle}
\usepackage[latin1]{inputenc}
\usepackage{multirow}
\usepackage{young}

\def\1{\mbox{l\hspace{-0.53em}1}}
\begin{document}
\title{The wave function of $[\bf{70},1^-]$ baryons in the $1/N_c$ expansion}

\author{N. Matagne\footnote{present address: Institut f\"ur Theoretische Physik, Universit\"at Giessen, D-35392  Giessen, Germany \\ e-mail address: Nicolas.Matagne@theo.physik.uni-giessen.de}}

\author{Fl. Stancu\footnote{e-mail address: fstancu@ulg.ac.be
}}
\affiliation{University of Li\`ege, Institute of Physics B5, Sart Tilman,
B-4000 Li\`ege 1, Belgium}

\date{\today}

\begin{abstract}
\baselineskip=0.50cm
Much work has been devoted to the study of nonstrange baryons 
belonging to the  $[\textbf{70},1^-]$ multiplet in the framework 
of the $1/N_c$ expansion.
Using group theoretical arguments here we examine
the relation between the exact wave function and the approximate one, 
customarily used in applications where
the system is separated into a ground state core and an excited quark. 
We show that the exact and approximate  wave functions globally give 
similar results for all of mass operators presented in this work.
However we find that the inclusion of operators acting 
separately on the core and on the excited quark deteriorates the fit
and leads to unsatisfactory values for the coefficients which encode
the quark dynamics. 
Much better results are obtained when we
include operators acting on the whole system, both for the exact
and the approximate wave function. 
\end{abstract}

\maketitle
\section{Introduction}
The $1/N_c$ expansion of QCD \cite{tHo74,Wit79,GS84} 
is an interesting and systematic approach 
to study baryon spectroscopy. It has been applied to the ground state
baryons \cite{DM93,DJM94,DJM95,CGO94,Jenk1,JL95,DDJM96} as well as to excited
states, in particular to the negative parity  
$[\textbf{70},1^-]$ multiplet 
($N = 1$ band)
\cite{Goi97,PY,CCGL,CGKM,CaCa98,SGS}, to the positive parity Roper resonance
$[\textbf{56'},0^+]$  ($N = 2$ band) \cite{CC00}, to the  
$[\textbf{56},2^+]$  \cite{GSS03}
and  the $[\textbf{70},\ell^+]$ multiplets ($\ell$ = 0 and 2) \cite{MS2,Matagne:2006zf}, both belonging to
the $N = 2$ band
and to the $[\textbf{56},4^+]$ multiplet ($N = 4$ band) \cite{MS1}.
Estimates for the lowest multiplet $[\textbf{70},3^-]$ 
of the $N = 3$ band have also been
made \cite{GM}. 
In this approach 
the main features of the constituent quark model emerge naturally
\cite{SBMS,SBS}
as for example the dominant role of the spin-spin term and the
smallness of the spin-orbit term.

The study of excited states belonging to the symmetric representation
$[\textbf{56}]$ is similar to that of the ground state. The introduction
of an orbital part in the wave function does not affect 
the procedure, inasmuch as the flavor-spin part remains symmetric.
The study of excited states which are mixed symmetric, 
both in the orbital and flavor-spin space is more complicated and it 
became controversial. 
Presently we discuss  the $[\textbf{70},1^-]$ multiplet, which is 
the simplest one, having 7 experimentally known resonances 
\cite{PDG04} in the nonstrange sector.
The standard procedure \cite{CCGL} is to separate the system into 
of a ground state core of $N_c-1$ quarks  and an 
orbitally excited $(\ell = 1)$ quark. Then, in the spirit of a Hartree picture 
the system is described by an approximate wave function, where the orbital
part has a configuration of type $s^{N_c-1} p$ (no antisymmetrization) 
which is combined with an approximate spin-flavor part. 
There is also a more straightforward procedure where the
system is treated as a whole  \cite{NEWLOOK}. This procedure is in
the spirit of the {\it spectrum generating algebra} method, introduced 
by  Gell-Mann and Ne'eman. It has been applied to
the ground state,  where use of the generators describing
the entire system is made. In this procedure the exact wave function is needed.

Here we reanalyze  the $[\textbf{70},1^-]$ 
multiplet ($N = 1$ band) 
without any prejudice. 
We start with the standard procedure based on the core+excited quark separation.
We show the relation between the approximate  
wave function and the exact one, constructed explicitly in the next section. 
To test the validity of the approximate wave function \cite{CCGL}
we first compare the analytical expressions of the matrix elements
of various operators entering the mass formula. Next we perform a numerical 
fit to the data. For practical reasons we choose a simple mass operator
containing the most dominant terms in order  
to fit the 7 experimentally known masses and ultimately the two mixing angles
also.  

The next section is devoted to the analysis of the wave function and its 
construction by using isoscalar factors. In Sec. III we briefly introduce
the mass operator. In Sec. IV we perform  fits to the data, by 
including various sets of invariant operators constructed from
the generators of SU(4) and O(3). Sec. V is devoted to mixing angles. The last 
section contains our 
conclusions. Appendix A describes the procedure used to construct
the needed isoscalar factors of the permutation group. Appendix B
introduces the fractional parentage coefficients needed to separate
the orbital wave function into a part describing the core formed of $N_c-1$ 
quarks and a single quark.
Appendix C provides analytic expressions of the matrix elements
necessary to calculate the isoscalar factors.    
\section{The exact wave function}

The system under concern has four degrees of freedom: orbital (O), 
flavor (F), 
spin (S) and color (C). It is then useful to construct its wave function 
with the help 
of inner products of $S_{N_c}$ in order to fulfill the Fermi statistics.
The color part being antisymmetric, the orbital-spin-flavor 
part must be symmetric. As in the mass formula there are no
color operators, the color part being integrated out, we are concerned
with the orbital-spin-flavor part only. 
 
A convenient way to construct a wave function of given symmetry $[f]$ is
to use the inner product $[f'] \times [f'']$ of two irreducible 
representations (irreps) of $S_{N_c}$, generally reducible into a 
Clebsch-Gordan series with a given number of irreps $[f]$. 
A basis vector of $[f]$ is denoted by $|[f]Y \rangle$, where $Y$ is
the corresponding Young tableau (or Yamanouchi symbol). Its 
general form is \cite{book}
\begin{equation}\label{TOT}
|[f]Y \rangle = \sum_{Y',Y''} S([f']Y'[f'']Y''|[f]Y) |[f']Y' \rangle |[f'']Y'' \rangle,
\end{equation}
where $S([f']Y'[f'']Y''|[f]Y)$ are Clebsch-Gordan (CG) coefficients 
of $S_{N_c}$.

Here we deal with a system of $N_c$ quarks having one unit of
orbital excitation.  Hence
the orbital part must have a mixed symmetry $[N_c-1,1]$. To get a totally 
symmetric state $|[N_c]1 \rangle$, where for brevity the Yamanouchi symbol 
1...1 ($N_c$ times) is
denoted by 1, the FS part must belong to the irrep  $[N_c-1,1]$ as well.
Then the wave function takes the form
\begin{equation}
\label{EWF}
|[N_c]1 \rangle = {\left(\frac{1}{N_c-1}\right)}^{1/2}
\sum_{Y} |[N_c-1,1] Y \rangle_O  |[N_c-1,1] Y \rangle_{FS},
\end{equation}    
where the coefficient in front is the CG coefficient needed to construct
a symmetric state of $N_c$ quarks. 
The sum is performed over all possible $N_c - 1$ standard Young tableaux. 
In this sum there is only one $Y$ (the normal Young tableau) 
where the last particle is in 
the second row and $N_c-2$ terms with the $N_c$-th particle in the 
first row. 
If we specify 
the row $p$ of the $N_c$-th quark and the row $q$ of the $(N_c-1)$-th quark, 
and denote by $y$ the distribution of the $N_c-2$ remaining quarks, we can 
write $Y$ more explicitly as
\begin{equation}
Y = (pqy).
\end{equation} 
For illustration let us consider the case $N_c$ = 5. Expressed in terms of Young 
tableaux, the wave function 
(\ref{EWF}) reads
\begin{eqnarray}
\lefteqn{\raisebox{-3.5pt}{\mbox{\begin{Young}
      1 & 2 & 3 & 4 & 5 \cr
      \end{Young}}}
 = } \nonumber \\ & &  \frac{1}{\sqrt{4}} \left[\left(\negthinspace \negthinspace\negthinspace\negthinspace
\raisebox{-10.0pt}{\mbox{\begin{Young}
1 & 2 & 3 & 4 \cr
5 \cr
\end{Young}}}\;\right)_{\mathrm{O}}
\left(\negthinspace \negthinspace\negthinspace\negthinspace\raisebox{-10.0pt}{\mbox{\begin{Young}
1 & 2 & 3 & 4\cr
5 \cr
\end{Young}}}\;\right)_{\mathrm{FS}} 
+ \left(\negthinspace\negthinspace \negthinspace\negthinspace\raisebox{-10.pt}{\mbox{\begin{Young}
1 & 2 & 3 & 5 \cr
4 \cr
\end{Young}}}\;\right)_{\mathrm{O}}
\left(\negthinspace\negthinspace \negthinspace\negthinspace\raisebox{-10.0pt}{\mbox{\begin{Young}
1 & 2 & 3 & 5 \cr
 4 \cr
\end{Young}}}\;\right)_{\mathrm{FS}}
\right. \nonumber \\
& & +  \left. \left(\negthinspace \negthinspace \negthinspace\negthinspace\raisebox{-10.0pt}{\mbox{\begin{Young}
1 & 2 & 4 & 5 \cr
3 \cr
\end{Young}}}\;\right)_{\mathrm{O}}
\left(\negthinspace \negthinspace\negthinspace\negthinspace\raisebox{-10.0pt}{\mbox{\begin{Young}
1 & 2 & 4 & 5\cr
3 \cr
\end{Young}}}\;\right)_{\mathrm{FS}}
+ \left(\negthinspace \negthinspace\negthinspace\negthinspace\raisebox{-10.0pt}{\mbox{\begin{Young}
1 & 3 & 4 & 5 \cr
2 \cr
\end{Young}}}\;\right)_{\mathrm{O}}
\left(\negthinspace \negthinspace\negthinspace\negthinspace\raisebox{-10.0pt}{\mbox{\begin{Young}
1 & 3 & 4 & 5 \cr
 2 \cr
\end{Young}}}\;\right)_{\mathrm{FS}}
\right].
\label{EWF5}
\end{eqnarray}
In this notation, in the left hand side one has $p = q = 1$ because 
the $N_c$-th particle, \emph{i.e.} particle
5,  and the $(N_c-1)$-th, \emph{i.e.} particle 4, are both in the first row. 
The first term in the right hand side of (\ref{EWF5}) 
has $p = 2$ and $q = 1$ both for (O) and (FS) parts. This is the only 
term considered in  Ref. \cite{CCGL}. 
The second term has $p = 1$ and  $q = 2$ and the last two terms have 
$p = q = 1$. In the following we shall mostly concentrate
on the $p = 1$ terms in the right hand side, neglected in  Ref. \cite{CCGL} and compare results derived from the exact
wave function (\ref{EWF}) with those obtained in Ref. \cite{CCGL}. To 
include the neglected terms we have to explicitly consider the (FS) part
of the wave function, as follows.

At its turn, the (FS) part can be decomposed into a sum of products of
spin (S) and flavor (F) parts using Eq. (\ref{TOT})
where in the left hand side we now have 
$|[f]Y \rangle = |[N_c-1,1]Y \rangle_{FS}$ and in the right hand side the 
appropriate CG coefficients. For these coefficients
it is useful to use a factorization property
(Racah's {\it factorization lemma}) which helps to
decouple the $N_c$-th particle from the rest. 
Accordingly, every CG coefficient of  
$S_{N_c}$ can be factorized into an isoscalar factor times a CG 
coefficient of $S_{N_c-1}$, and so on. We apply this property to 
$S_{N_c} \supset S_{N_c-1}$ only. To do so, it is necessary to specify 
the row $p$ of the $N_c$-th and the row $q$ of the $(N_c-1)$-th quarks, 
as above. 
If the isoscalar factor is denoted by $K([f']p'[f'']p''|[f]p)$
the factorization property reads
\begin{equation}\label{racah}
S([f']p'q'y' [f'']p''q''y'' | [f]pqy ) =
K([f']p'[f'']p''|[f]p) S([f'_{p'}]q'y' [f''_{p''}]q''y'' | [f_p]qy ),
\end{equation}
where the second factor in the right hand side is a CG coefficient of
S$_{N_c-1}$ containing the 
partitions $[f'_{p'}]$, $[f''_{p''}]$ and $[f_p]$ obtained after 
the removal of the $N_c$-th quark.

In Ref.  \cite{MS2} we have derived the isoscalar factors associated to 
the normal Young tableau of $|[N_c-1,1]Y \rangle_{FS}$, \emph{i.e.} for $p = 2$,
$q = 1$. 
We have also shown  
that the coefficients  $c_{\rho \eta}$ appearing in the
wave function of Ref. \cite{CCGL}
correspond precisely to these isoscalar factors. In the present 
work we derive  the isoscalar factors of the other $N_c-2$ terms,
having $p = 1$, for the first time. Details are given in Appendices A and C.
The results are summarized in Tables I, II and III.
In each table the column corresponding to $p = 1$ is new.
For completeness, we also reproduce
the results of Ref. \cite{MS2} for $p = 2$.
They are both used in the calculation of matrix elements of various 
operators entering the mass formula. As shown below, the total wave function 
needed to calculate these matrix elements
contains a spin-flavor part, Eq. (\ref{fs}), constructed with the help of  
the isoscalar factors of Tables I, II and III.

The isoscalar factors derived here
can be applied to the study of other $[\textbf{70},\ell^P]$ multiplets
with $\ell \neq 1$ and parity $P = + $ or $P = -$,
or to other physical systems having $N$ fermions. They 
satisfy a recurrence relation 
described in Refs. \cite{book,ISOSC} which has been used to check
the presently derived analytic results in the case of $N_c$ = 5, 7 and 9 
quarks. The results from Tables I, II and III can also be
interpreted as isoscalar factors of SU(4), related to
CG coefficients appearing in the chain SU(4) $\supset$ SU$_S$(2) $\times$ 
SU$_I$(2) with the specific values of $S$ and $I$ presented here.
 
For the orbital part of the wave function (\ref{EWF}) one can use 
one-body fractional parentage coefficients (cfp) to decouple
the $N_c$-th quark from the rest. Together with 
the factorization property (\ref{racah}) it leads to a 
form of the wave function which is a sum of products 
of a wave function describing a system of $N_c-1$ quarks, called ``core'' times
the wave function of the $N_c$-th quark. If couplings are introduced
this gives 
\begin{equation}\label{product}
|\ell S J J_3; I I_3 \rangle = 
\sum_{p,\ell_c,\ell_q} a(p,\ell_c,\ell_q)~ |\ell_c \ell_q \ell S J J_3; I I_3 \rangle_p, 
\end{equation}
where $a(p,\ell_c,\ell_q)$ are the one-body cfp given in Appendix B,
with $\ell_c$ and $\ell_q$ representing the angular momentum
of the core and of the decoupled quark respectively. One can write
\begin{eqnarray}\label{wfp}
\lefteqn{|\ell_c \ell_q \ell S J J_3; I I_3 \rangle_p  = }\nonumber \\ & & 
\sum_{m_c,m_q,m_\ell,S_3}
\left(\begin{array}{cc|c}
	\ell_c    &  \ell_q   & \ell   \\
	m_c  &    m_q    & m_\ell 
      \end{array}\right) 
   \left(\begin{array}{cc|c}
	\ell    &    S   & J   \\
	m_\ell  &    S_3  & J_3 
      \end{array}\right)
|\ell_c m_c \rangle |\ell_q m_q \rangle       
|[N_c-1,1]p;S S_3; I I_3 \rangle,
\end{eqnarray}
containing the spin-flavor part 
\begin{eqnarray}\label{fs}
|[N_c-1,1]p;S S_3; I I_3 \rangle = 
\sum_{p' p''} K([f']p'[f'']p''|[N_c-1,1]p) 
|S S_3;p' \rangle |I I_3;p'' \rangle  
\end{eqnarray}
where 
\begin{equation}\label{spin}
|S S_3; p' \rangle = \sum_{m_1,m_2}
 \left(\begin{array}{cc|c}
	S_c    &    \frac{1}{2}   & S   \\
	m_1  &         m_2        & S_3
      \end{array}\right)
      |S_cm_1 \rangle |1/2m_2 \rangle,
\end{equation}
with $S_c = S - 1/2$ if $p' = 1$    and $S_c = S + 1/2$ if   $p' = 2$ and 
\begin{equation}\label{isospin}
|I I_3; p'' \rangle = \sum_{i_1,i_2}
 \left(\begin{array}{cc|c}
	I_c    &    \frac{1}{2}   & I   \\
	i_1    &       i_2        & I_3
      \end{array}\right)
      |I_c i_1 \rangle |1/2 i_2 \rangle,
\end{equation}
with $I_c = I - 1/2$ if $p'' = 1$   and   $I_c = I + 1/2$  if $p'' = 2$. Thus $p'$
and $p''$ represent here the position of the $N_c$-th quark in the spin and 
isospin parts of the wave function  
$|S S_3;p' \rangle |I I_3;p'' \rangle $
respectively. Note that the wave function (\ref{fs})
should also contain the second factor  
$S([f'_{p'}]q'y' [f''_{p''}]q''y'' | [f_p]qy )$
of Eq. (\ref{racah}) together with  sums over 
$q' y'$ and $q'' y''$ which ensures 
the proper permutation symmetry of the spin-flavor part of the core 
wave function 
after the removal of the $N_c$-th quark. 
However, these  CG coefficients of S$_{N_c-1}$ do not need to be known and 
for simplicity we ignore them because their squares  
add up together to 1, by the standard normalization of CG coefficients.
Equations (\ref{product})-(\ref{isospin}) are needed 
in the calculation of operators as $s \cdot S_c$, $S^2_c$, $t \cdot T_c $, $T^2_c$, etc.
(see next section) which involve the core explicitly. 

Let us illustrate, for example,  the use of  Table I, when $p = 1$. Then Eq. (\ref{fs}) becomes
\begin{eqnarray}
\lefteqn{\left|[N_c-1,1]p=1;\frac{1}{2} S_3;\frac{1}{2} I_3 \right\rangle =}\nonumber \\ & &\sqrt{\frac{N_c-3}{2(N_c-2)}} 
\left|\frac{1}{2} S_3; p'= 2 \right\rangle  \left|\frac{1}{2} I_3; p''= 2 \right\rangle \nonumber \\
& & - \sqrt{\frac{N_c-1}{4(N_c-2)}} \left|\frac{1}{2} S_3; p'= 2 \right\rangle  \left|\frac{1}{2} I_3; p''= 1 \right\rangle  
- \sqrt{\frac{N_c-1}{4(N_c-2)}} \left|\frac{1}{2} S_3; p'= 1 \right\rangle  \left|\frac{1}{2} I_3; p''= 2 \right\rangle,\nonumber \\ 
\end{eqnarray}
where $S_c$ and $I_c$ are determined as explained below Eqs. (\ref{spin}) and (\ref{isospin}).

\begin{table}
\renewcommand{\arraystretch}{1.8}
 \begin{tabular}{c|ccc}
$[f']p'[f'']p''$   & $[N_c-1,1]1$  &\hspace{3cm}& $[N_c-1,1]2$  \\ \hline
$\left[ \frac{N_c+1}{2},\frac{N_c-1}{2}\right ] 1\left\lbrack \frac{N_c+1}{2},\frac{N_c-1}{2}\right \rbrack 1$ &    0    &  & $-\sqrt{\frac{3(N_c-1)}{4N_c}}$ \\
$\left\lbrack \frac{N_c+1}{2},\frac{N_c-1}{2}\right \rbrack 2 \left\lbrack \frac{N_c+1}{2},\frac{N_c-1}{2}\right \rbrack 2$ & $\sqrt{\frac{N_c-3}{2(N_c-2)}}$ &\hspace{1cm} & $\sqrt{\frac{N_c+3}{4N_c}}$ \\
$\left\lbrack \frac{N_c+1}{2},\frac{N_c-1}{2}\right \rbrack 2 \left\lbrack \frac{N_c+1}{2},\frac{N_c-1}{2}\right \rbrack 1$ &   $-\frac{1}{2}\sqrt{\frac{N_c-1}{N_c-2}}$ & & 0 \\
$\left\lbrack \frac{N_c+1}{2},\frac{N_c-1}{2}\right \rbrack 1 \left\lbrack \frac{N_c+1}{2},\frac{N_c-1}{2}\right \rbrack 2$ &   $-\frac{1}{2}\sqrt{\frac{N_c-1}{N_c-2}}$ & & 0 \\
\hline
 \end{tabular}
\caption{Isoscalar factors $K([f']p'[f'']p''| [f]p)$ for
$S=I=1/2$, corresponding to $^{2}8$ when $N_c=3$. The second column gives
results for $p = 1$ and the third for $p = 2$.}\label{spin1/2n}
\end{table}

\begin{table}
\renewcommand{\arraystretch}{1.8}
 \begin{tabular}{c|ccc}
$[f']p'[f'']p''$   & $[N_c-1,1]1$  &\hspace{3cm}& $[N_c-1,1]2$  \\ \hline
$\left\lbrack \frac{N_c+3}{2},\frac{N_c-3}{2}\right \rbrack 1 \left\lbrack \frac{N_c+1}{2},\frac{N_c-1}{2}\right \rbrack 1$ & $\frac{1}{2}\sqrt{\frac{(N_c-1)(N_c+3)}{N_c(N_c-2)}}$ & \hspace{1cm} & 0 \\
$\left\lbrack \frac{N_c+3}{2},\frac{N_c-3}{2}\right \rbrack 2 \left\lbrack \frac{N_c+1}{2},\frac{N_c-1}{2}\right \rbrack 2$ & $\frac{1}{2}\sqrt{\frac{5(N_c-1)(N_c-3)}{2N_c(N_c-2)}}$ & \hspace{1cm} & 0 \\
$\left\lbrack \frac{N_c+3}{2},\frac{N_c-3}{2}\right \rbrack 1 \left\lbrack \frac{N_c+1}{2},\frac{N_c-1}{2}\right \rbrack 2$ & $\frac{1}{2}\sqrt{\frac{(N_c-3)(N_c+3)}{2N_c(N_c-2)}}$ & & 1 \\
 $\left\lbrack \frac{N_c+3}{2},\frac{N_c-3}{2}\right \rbrack 2 \left\lbrack \frac{N_c+1}{2},\frac{N_c-1}{2}\right \rbrack 1$ & 0 & & 0 \\
\hline
 \end{tabular}
\caption{Isoscalar factors $K([f']p'[f'']p''| [f]p)$ for
$ S=3/2,\ I=1/2$, corresponding to $^{4}8$ when $N_c=3$. The second column gives
results for $p = 1$ and the third for $p = 2$.}\label{spin3/2n}
\end{table}

In the following we shall 
compare results obtained from the exact wave function (\ref{product})
with those obtained from a truncated wave function, representing only 
the term with $p = 2$ in the sum (\ref{product}). In addition, by taking 
$a(2,\ell_c = 0, \ell_q = 1) = 1 $, instead of the value given
in Appendix B, and renormalizing the resulting wave function, we recover the 
wave function of Ref. \cite{CCGL}. As seen from Appendix B, it is entirely reasonable to take   
$a(2,\ell_c = 0, \ell_q = 1) = 1$  when $N_c$ is large. Then the 
core is in its ground state. It is in fact 
the validity  of this wave function, representing the Hartree approximation, 
we wish to analyze as compared to the exact wave function in the
description of the $[{\bf 70},1^-]$ multiplet.

\begin{table}
\renewcommand{\arraystretch}{1.8}
 \begin{tabular}{c|ccc}
$[f']p'[f'']p''$   & $[N_c-1,1]1$  &\hspace{3cm}& $[N_c-1,1]2$  \\ \hline
$\left\lbrack \frac{N_c+1}{2},\frac{N_c-1}{2}\right \rbrack 1 \left\lbrack \frac{N_c+3}{2},\frac{N_c-3}{2}\right
\rbrack 1$& $\frac{1}{2}\sqrt{\frac{(N_c-1)(N_c+3)}{N_c(N_c-2)}}$ & \hspace{1cm} & 0 \\
$\left\lbrack \frac{N_c+1}{2},\frac{N_c-1}{2}\right \rbrack 2 \left\lbrack \frac{N_c+3}{2},\frac{N_c-3}{2}\right \rbrack 2$ & $\frac{1}{2}\sqrt{\frac{5(N_c-1)(N_c-3)}{2N_c(N_c-2)}}$ & \hspace{1cm} & 0 \\
$\left\lbrack \frac{N_c+1}{2},\frac{N_c-1}{2}\right \rbrack 2 \left\lbrack \frac{N_c+3}{2},\frac{N_c-3}{2}\right \rbrack 1$ & $\frac{1}{2}\sqrt{\frac{(N_c-3)(N_c+3)}{2N_c(N_c-2)}}$ & & 1 \\
$\left\lbrack \frac{N_c+1}{2},\frac{N_c-1}{2}\right \rbrack 1 \left\lbrack
\frac{N_c+3}{2},\frac{N_c-3}{2}\right \rbrack 2$ & 0 & & 0 \\
\hline
 \end{tabular}
\caption{Isoscalar factors $K([f']p'[f'']p''| [f]p)$ for
$S=1/2,\ I=3/2$, corresponding to $^{2}10$ when $N_c=3$. The second column gives
results for $p = 1$ and the third for $p = 2$.}\label{spin1/2delta}
\end{table}

\section{The mass operator}
The mass operator $M$ is a linear combination of independent operators $O_i$ 
\begin{equation}
\label{massoperator}
M = \sum_{i} c_i O_i,
\end{equation} 
where the coefficients $c_i$ are reduced matrix elements that
encode the QCD dynamics and are  
determined from a fit to the existing data. 
The operators $O_i$ are constructed from the  
SU(4) generators $S_i$, $T_a$ and $G_{ia}$ and the 
SO(3) generators $\ell_i$. For the purpose of the present 
analysis it is enough to restrict the choice of the operators 
$O_i$ to a few selected dominant operators. 
Our previous analysis \cite{NEWLOOK} suggests that the dominant 
operators up to order $\mathcal{O}({1/N_c})$ included are those constructed 
from the SU(4) generators exclusively, the operators containing
angular momentum having a minor role. Among them, 
the spin-orbit, although
weak, is however considered here. Samples are given in Tables VI-X.
(For convenience and clarity we denote  some operators in the
sum (\ref{massoperator})   by $O'_i$. The meaning will be obvious later on.)

In Tables IV and V we present the analytic expressions of the matrix 
elements of the operators containing spin and isospin, used in this study. 
They have been obtained either with
the approximate or with the exact wave function. One can see that
the analytic forms are different in the two cases. This implies that
at $N_c = 3$ the values obtained for $c_i$ from the fit to data
are expected to be also different, as illustrated in the next section. 
The matrix elements of the spin-orbit operator are identical for
the exact and an approximate wave function and can be found 
in Table II of Ref. \cite{CCGL}.
 
\begin{table}
\renewcommand{\arraystretch}{1.25}
\begin{tabular}{c||c|c|c|c|} \cline{2-5}
 & \multicolumn{2}{c|}{$\langle s\cdot S_c\rangle$} & \multicolumn{2}{c|}{$\langle S^2_c\rangle$}\\ \cline{2-5}
& approx. w.f. & exact w.f. & approx. w.f. & exact w.f. \\ \hline
$^2 8$ & $ -\frac{N_c+3}{4N_c}$ & $-\frac{3(N_c-1)}{4N_c}$ & $\frac{N_c+3}{2N_c}$  & $\frac{3(N_c-1)}{2N_c}$ \\ 
$^4 8$ & $\frac{1}{2}$  & $-\frac{3(N_c-5)}{4N_c}$  & 2 & $\frac{3(3N_c-5)}{2N_c}$  \\ 
$^2 10$ &  $-1$ & $-\frac{3(N_c-1)}{4N_c}$& 2 &  $\frac{3(N_c-1)}{2N_c}$ \\ \hline
\end{tabular}
\caption{Matrix elements of the spin operators calculated with the 
approximate and the exact wave functions.}
\end{table}

\begin{table}
\renewcommand{\arraystretch}{1.25}
\begin{tabular}{c||c|c|c|c|} \cline{2-5}
 & \multicolumn{2}{c|}{$\langle t\cdot T_c\rangle$} & \multicolumn{2}{c|}{$\langle T^2_c\rangle$}\\ \cline{2-5}
& approx. w.f. & exact w.f. & approx. w.f. & exact w.f. \\ \hline
$^2 8$ & $ -\frac{N_c+3}{4N_c}$ & $-\frac{3(N_c-1)}{4N_c}$ & $\frac{N_c+3}{2N_c}$  & $\frac{3(N_c-1)}{2N_c}$ \\ 
$^4 8$ &  $-1$ & $-\frac{3(N_c-1)}{4N_c}$& 2 &  $\frac{3(N_c-1)}{2N_c}$ \\ 
$^2 10$ & $\frac{1}{2}$  & $-\frac{3(N_c-5)}{4N_c}$  & 2 & $\frac{3(3N_c-5)}{2N_c}$  \\ \hline
\end{tabular}
\caption{Matrix elements of the isospin operators calculated with the 
approximate and the exact wave functions.}
\end{table}

\section{Fit and discussion}

In the numerical fit we have considered the masses of the seven experimentally
known nonstrange resonances belonging to the
$[\textbf{70},1^-]$ multiplet. Neglecting the mixing $^4 N -\; \negthinspace\negthinspace ^2N$ we 
identify them as:
$^2 N_{1/2}(1538 \pm 18)$, $^4 N_{1/2}(1660 \pm 20)$,
$^2 N_{3/2}(1523 \pm 8)$,  $^4 N_{3/2}(1700 \pm 50)$,
$^4 N_{5/2}(1678 \pm 8)$,  $^2 \Delta_{1/2}(1645 \pm 30)$ and
$^2 \Delta_{3/2}(1720 \pm 50)$.

The results from various fits are presented in Tables VI-IX.
In Table VI the first four operators are among those considered 
in Ref. \cite{CCGL}, where the system is decoupled into a ground
state core and an excited quark.  To them we have added the isospin operator
which we expect to play an important role. The value of $\chi_{\mathrm{dof}}^2$
is satisfactory but the value of the coefficient $c_1$ is 
much smaller than in previous studies where it
generally appears to be of the order of 500 MeV.  
The spin-orbit
coefficient is small, as expected, but the coefficients $c_3$ and $c_4$  
are exceedingly large in absolute values and have opposite signs, which
suggest some compensation.
The coefficient $c_5$ is also very large and negative for the 
approximate wave function but it has a reasonable value for 
the exact wave function.
A solution is either to eliminate the isospin operator,
but this cannot be theoretically  justified or to make some combinations 
of the above operators. In Table VII we consider the linear combination
$ 2 s^iS_c^i+S_c^iS_c^i+\frac{3}{4} = S^2, $ \emph{i.e.} we use the total spin
operator of the system, entirely legitimate in the $1/N_c$ expansion
approach. The coefficient $c_1$ acquires the expected value
and the spin coefficient $c'_3$ has
a reasonable value, being about 70 MeV smaller for the
exact wave function  than for the approximate one. 
   The isospin coefficient is equal in both 
cases and also has a reasonable value. By analogy, in Table VIII  we include the 
linear combination
$ 2 t^aT_c^a+T_c^aT_c^a+\frac{3}{4} = I^2 $ \emph{i.e.} we use the total 
isospin and restrict the spin contribution to  $\frac{1}{N_c}s \cdot S_c$.
Then the isospin coefficient $c'_5$ acquires  values
comparable to the spin contribution in Table VII. The spin coefficient
$c_3$ has now values comparable to the isospin  coefficient $c_5$ in Table 
VII. One can infer that the spin and isospin  play a similar role
in the mass formula. 
This statement is clearly proved in Table IX where we include the spin and
isospin contributions on an equal footing, \emph{i.e.} we include operators
proportional to  $S^2$ and $I^2 $. The situation is entirely identical
for the exact and approximate wave function, with reasonable values 
for all coefficients and a  $\chi_{\mathrm{dof}}^2$ = 1.04. The identity  
of the results is natural because both the approximate and the exact wave function
describe a system  of a given spin $S$ and isospin $I$.  

\begin{table}
\renewcommand{\arraystretch}{1.15}
 \begin{tabular}{lllclc}
\hline \hline
Operator  &   \hspace{1cm} &  &    Approx. w.f. (MeV) & \hspace{1cm} & Exact w.f. (MeV) \\ \hline
$O_1 = N_c \ \1 $    &  &  $c_1 =  $  & $211 \pm 23$ & & $299 \pm 20$ \\
$O_2 = \ell^i s^i$    &  & $c_2 = $ & $3 \pm 15$ & & $3 \pm 15$ \\
$O_3 = \frac{1}{N_c} s^iS_c^i $ & & $c_3= $ & $-1486 \pm 141$ & & $-1096\pm 125$\\
$O_4 = \frac{1}{N_c} S_c^iS_c^i$ & & $c_4 = $ & $1182\pm 74$ & & $1545 \pm 122$ \\
$O_5 = \frac{1}{N_c} t^aT_c^a$ & & $c_5 = $ & $-1508\pm 149$  & & $417 \pm 79$ \\ 
\hline
$\chi_{\mathrm{dof}}^2$       &    & & $1.56$ & & $1.56$      \\ \hline \hline 
\end{tabular}
\caption{List of operators and coefficients obtained in the numerical fit 
to the 7 known experimental masses of the lowest negative parity
resonances (see text).  For the operators 
$O_3$, $O_4$ and $O_5$ we use the matrix elements from Tables IV and V.}
\end{table}

\begin{table}
\renewcommand{\arraystretch}{1.15}
 \begin{tabular}{lllclc}
\hline \hline
Operator  &   \hspace{1cm} &  &    Approx. w.f. (MeV) & \hspace{1cm} & Exact w.f. (MeV) \\ \hline
$O_1 = N_c \ \1 $    &  &  $c_1 =  $  & $513 \pm 4$ & & $519 \pm 5$ \\
$O_2 = \ell^i s^i$    &  & $c_2 = $ & $3 \pm 15$ & & $3 \pm 15$ \\
$O'_3 = \frac{1}{N_c}\left(2 s^iS_c^i+S_c^iS_c^i+\frac{3}{4}\right)$ & & $c'_3 = $ & $219\pm 19$  & & $150\pm 11$ \\
$O_5 = \frac{1}{N_c} t^aT_c^a$ & & $c_5 = $ & $417\pm 80$ & & $417\pm 80$ \\
\hline
$\chi_{\mathrm{dof}}^2$       &    & & $1.04$ & & $1.04$      \\ \hline \hline 
\end{tabular}
\caption{Same as Table VI but for $O'_3$, which combines  $O_3$ and $O_4$
instead of using them separately. }
\end{table}

\begin{table}
\renewcommand{\arraystretch}{1.15}
 \begin{tabular}{lllclc}
\hline \hline
Operator  &   \hspace{1cm} &  &    Approx. w.f. (MeV) & \hspace{1cm} & Exact w.f. (MeV) \\ \hline
$O_1 = N_c \ \1 $    &  &  $c_1 =  $  & $516 \pm 3$ & & $522 \pm 3$ \\
$O_2 = \ell^i s^i$    &  & $c_2 = $ & $3 \pm 15$ & & $3 \pm 15$ \\
$O_3 = \frac{1}{N_c}s^iS_c^i$ & & $c_3 = $ & $450 \pm 33$ & & $450 \pm 33$ \\
$O'_5 = \frac{1}{N_c}\left(2 t^aT_c^a+T_c^aT_c^a+\frac{3}{4}\right)$ & &$c'_5 = $ & $214\pm 28$ & & $139 \pm 27$ \\ 
\hline
$\chi_{\mathrm{dof}}^2$       &    & & $1.04$ & & $1.04$      \\ \hline \hline 
\end{tabular}
\caption{Same as Table VII but combining isospin operators instead of spin operators. }
\end{table}

\begin{table}
\renewcommand{\arraystretch}{1.15}
 \begin{tabular}{lllclc}
\hline \hline
Operator  &   \hspace{1cm} &  &    Approx. w.f. (MeV) & \hspace{1cm} & Exact w.f. (MeV) \\ \hline
$O_1 = N_c \ \1 $    &  &  $c_1 =  $  & $484 \pm 4$ & & $484 \pm 4$ \\
$O_2 = \ell^i s^i$    &  & $c_2 = $ & $3 \pm 15$ & & $3 \pm 15$\\
$O'_3 = \frac{1}{N_c}\left(2 s^iS_c^i+S_c^iS_c^i+\frac{3}{4}\right)$ & & $c'_3 = $ & $150 \pm 11$ & & $150 \pm 11$\\
$O'_5 = \frac{1}{N_c}\left(2 t^aT_c^a+T_c^aT_c^a+\frac{3}{4}\right) $ & &$c'_5= $ & $139 \pm 27$ & & $139 \pm 27$  \\ \hline
$\chi_{\mathrm{dof}}^2$       &    & & $1.04$ & & $1.04$ \\ \hline \hline
\end{tabular}
\caption{Same as Table VI but with operators proportional to the SU(2)-spin 
and SU(2)-isospin Casimir operators (see text). }
\end{table}

\section{Mixing angles}

So far we have discussed the mass spectrum. Additional empirical 
information come from the mixing angles extracted from the electromagnetic
and strong decays, in particular from the dominance 
of $N \eta$ decay of $N_{1/2}(1538 \pm 18)$. These angles are 
defined as
\begin{eqnarray}
|N_J(\mathrm{upper}) \rangle = \cos \theta_J |^4N_J \rangle +
 \sin \theta_J |^2N_J \rangle, \nonumber \\
|N_J(\mathrm{lower}) \rangle = \cos \theta_J |^2N_J \rangle -
 \sin \theta_J |^4N_J \rangle. 
\end{eqnarray}
Experimentally one finds 
$\theta^{exp}_{1/2} \approx   - 0.56$ rad and
$\theta^{exp}_{3/2} \approx  0.10$ rad  \cite{ISGUR}. 
In our case, the only operator with non-vanishing
matrix elements $^4 N -\; \negthinspace\negthinspace ^2N$, thus contributing to the mixing, 
is the  spin-orbit operator. Its off-diagonal matrix elements are
$\langle ^4 N_{1/2} | \ell \cdot s |  ^2N_{1/2} \rangle 
= - 1/3 $
and $\langle ^4 N_{3/2} | \ell \cdot s |^2N_{3/2} \rangle  
= - \sqrt{5/18} $, compatible with Ref. \cite{CCGL} when $N_c=3$.
In this notation the physical masses become
\begin{eqnarray}\label{MIXING}
M_J(\mathrm{upper}) = M(^4 N_{J}) 
+ c_2 \langle ^4 N_J | \ell \cdot s |  ^2N_J \rangle \tan \theta_J,
\nonumber\\
M_J(\mathrm{lower}) = M(^2 N_{J}) 
- c_2 \langle ^4 N_J | \ell \cdot s |  ^2N_J \rangle \tan \theta_J.
\end{eqnarray}
Due to the fact that the diagonal contribution of $O_2 = \ell \cdot s$
is very small, we expect its contribution
to the mixing to be small as well. By including 
the experimental values of $\theta_J$ in Eqs. (\ref{MIXING})
and applying again the minimization procedure,
we found indeed that the effect of mixing is negligible. The only
coefficient being slightly affected is $c_2$ and the 
$\chi^2_{\mathrm{dof}}$ remains practically the same.
Next we varied  $\theta_J$
to see if the value of $\chi_{\mathrm{dof}}^2$ 
remains stable. Indeed it does, for variations of $\theta_J$ of $\pm  0.05$ rad.
In Table X we show the optimal set of coefficients $c_i$
associated to the experimental values of $\theta_J$.
\begin{table}[ht]
\renewcommand{\arraystretch}{1.15}
 \begin{tabular}{lllclc}
\hline \hline
Operator  &   \hspace{1cm} &  &    Approx. w.f. (MeV) & \hspace{1cm} & Exact w.f. (MeV) \\ \hline
$O_1 = N_c \ \1 $    &  &  $c_1 =  $  & $484 \pm 4$ & & $484 \pm 4$ \\
$O_2 = \ell^i s^i$    &  & $c_2 = $ & $ -9\pm 15 $ & & $ -9\pm 15 $\\
$O'_3 = \frac{1}{N_c}\left(2 s^iS_c^i+S_c^iS_c^i+\frac{3}{4}\right)$ & & $c'_3 = $ & $150 \pm 11$ & & $150 \pm 11$\\
$O'_5 = \frac{1}{N_c}\left(2 t^aT_c^a+T_c^aT_c^a+\frac{3}{4}\right) $ & &$c'_5= $ & $139 \pm 27$ & & $139 \pm 27$  \\ \hline
$\chi_{\mathrm{dof}}^2$       &    & & $1.05$ & & $1.05$ \\ \hline \hline
\end{tabular}
\caption{Fit with state mixing for $N_{1/2}$ and $N_{3/2}$ due to the
spin-orbit coupling $\ell\cdot s$. }
\end{table}
In a more extended analysis as that of Ref. \cite{NEWLOOK} the operators 
$O_5 = \frac{15}{N_c}\ell^{(2)ij}G^{ia}G^{ja}$ and
$O_6 = \frac{3}{N_c}\ell^iT^aG^{ia}$  would also contribute.  
However all these operators contain angular momentum while
in constituent quark models where the 
predictions are generally good,  the source of mixing are either 
the spin-spin or the tensor interactions (no angular momentum),
contrary to the present case. We therefore believe that this 
important issue deserves further work in the future,
combined with a strong decay analysis, like in Ref. \cite{SGM}.

\section{Conclusions}

This study shows that the exact and the approximate wave function give
identical $\chi_{\mathrm{dof}}^2$ in this simplified fit and
rather similar results for the dynamical coefficients $c_i$ entering in
the mass formula. It also shows that the 
contribution of the isospin operator $O'_5 = \frac{1}{N_c} I^2$
is as important as that of the spin operator $O'_3 = \frac{1}{N_c} S^2$. 
In addition, it turned out that the separation of $S^2$ and of $I^2$ into independent
parts containing  core and excited quark operators is undesirable because it 
seriously deteriorates the fit.   The only satisfactory way to 
describe the spectrum is to include in  Eq. (\ref{massoperator})
the Casimir operators  of SU$_S$(2)  and SU$_I$(2) acting on the entire system. 

The difficulty with the approximate wave
function is that it cannot distinguish between $S^2_c$ and   $I^2_c$,
which act on the core only. In practice it means that the contribution 
of  $\frac{1}{N_c}I^2_c$ to the mass is hidden in the coefficient $c_i$ associated
to  $\frac{1}{N_c}S^2_c$. This also happens for $\frac{1}{N_c}I^2$ for states 
described by a symmetric flavor-spin
wave function 
where its contribution is absorbed by the coefficient of
$\frac{1}{N_c}S^2$. In other words  $S^2$ and $I^2$
should share their contribution to the mass by about the same amount.

In the decoupling scheme the isospin
can be introduced only through the operator $\frac{1}{N_c} t \cdot T_c$
which manifestly deteriorates the fit,  as seen in Table VI. This may
explain why this operator  has been avoided in 
numerical fits of previous studies  based on
core + quark separation \cite{CCGL}. 

While preserving the $S_{N_c}$ symmetry
in these calculations, our conclusion is 
at variance with that of Pirjol and Schat \cite{PISC} who formally claim that 
the inclusion of core and excited quark operators is necessary, 
as a consequence of constraints on the mass operator, resulting 
from the same $S_{N_c}$ symmetry.

The dynamical coefficients $c_i$ encode the quark dynamics in hadrons.
Therefore, to find their correct values and their evolution with 
the excitation energy is a very important task. 
They can serve to test the validity of quark models and quantitatively
determine the contribution of gluon and Goldstone boson exchange interactions,
both justified by the two-scale picture of Manohar and Georgi \cite{MG}.

As a byproduct we have obtained isoscalar factors of the permutation
group which may be useful in further studies of mixed symmetric
multiplets within the $1/N_c$
expansion or in any other approach where one fermion is separated
from the rest. 


{\bf Acknowledgments}.
The work of one of us (N. M.) was supported by the Institut Interuniversitaire 
des Sciences Nucl\'eaires (Belgium).

\par

\appendix
\section{Isoscalar factors}
Here we shortly describe the derivation of the isoscalar factors 
of Tables I,II and III.
We first recall that the expressions of $K([f']p'[f'']p''|[N_c-1,1]p)$
with $p = 2$ were obtained in Ref. \cite{MS2}.
We are presently concerned with the calculation of $K([f']p'[f'']p''|[N_c-1,1]p)$
for $p = 1$.

In the derivation below we use the symmetry property
\begin{equation}\label{symm}
K([f']p'[f'']p''|[f]p) = K([f'']p''[f']p'|[f]p),
\end{equation}
and the orthogonality relation
\begin{eqnarray}\label{orthog}
\sum_{p'p''}  K([f']p'[f'']p''|[f]p) K([f']p'[f'']p''|[f_1]p_1) & = &\delta_{f f_1}
\delta{p p_1}. \label{K1} 
\end{eqnarray}

\subsection{The case S = I = 1/2} 
In this case we have
\begin{equation}\label{partition}
[f'] = [f'']= \left[\frac{N_c+1}{2},\frac{N_c-1}{2}\right],
\end{equation}
and there are three nonzero coefficients necessary to express 
the FS state $|[N_c-1,1]1 \rangle$, as linear combinations of S and F
states. These are $ K([f']2[f'']2|[N_c-1,1]1)$, $K([f']1[f'']2|[N_c-1,1]1)$ and
$ K([f']2[f'']1|[N_c-1,1]1)$. But due to (\ref{symm}) the latter two are equal.
Then we are left with two unknown isoscalar factors which following 
(\ref{orthog}) are normalized as
\begin{equation}\label{norm}
(K([f']2[f'']2|[N_c-1,1]1))^2 + 2  (K([f']1[f'']2|[N_c-1,1]1))^2 = 1.
\end{equation}

To find the solution we need one more equation. This is provided
by the matrix elements of the operator $G^{ia}$ written as
\begin{equation}
\langle G^{ia} \rangle = \langle G^{ia}_c \rangle + \langle g^{ia}\rangle \label{sumg}
\end{equation}
with $G_c^{ia}$ and $g^{ia}$ acting on the core  and on the 
excited quark respectively.
According to Eqs. (\ref{GENsu4}) and (\ref{CASIMIR2}) the matrix elements of the SU(4) irrep 
$[N_c-1,1]$ with $S = I = 1/2$ are
\begin{eqnarray}
\hspace{-1cm}\left \langle \frac{1}{2} S'_3; \frac{1}{2} I'_3 \left|  
G^{ia} \right| \frac{1}{2} S_3; \frac{1}{2} I_3 \right\rangle  = & &   \nonumber \\
 & & \hspace{-2cm} -\frac{N_c - 6}{12} 
\left(\begin{array}{cc|c}
	1/2   &  1   &  1/2   \\
	S_3   &  i   &  S'_3 
      \end{array}\right) 
\left(\begin{array}{cc|c}
	1/2    &  1   &  1/2   \\
	I_3    &  a   &  I'_3 
      \end{array}\right).  
\label{G}
\end{eqnarray}
 
Taking $p = 1$ in  Eqs. (\ref{Gc}) and (\ref{g}) one obtains for 
$\langle G^{ia}_c \rangle$ and $\langle g^{ia}\rangle$ the following expressions
\begin{eqnarray}
\lefteqn{\left\langle [N_c-1,1]1; \frac{1}{2} S'_{3}; \frac{1}{2} I'_{3} \left| G^{ia}_c \right| 
[N_c-1,1]1; \frac{1}{2} S'_{3}; \frac{1}{2} I'_{3}\right\rangle
 =}\nonumber \\ & & \frac{1}{6}\left[\frac{N_c+3}{2}\left(K([f']2[f'']2|[N_c-1,1]1)\right)^2 +(N_c+3)\left(K([f']1[f'']2|[N_c-1,1]1)\right)^2 \right. \nonumber \\
& & \left. +\ 4 \sqrt{\frac{(N_c-3)(N_c-1)}{2}}K([f']2[f'']2|[N_c-1,1]1)K([f']1[f'']2|[N_c-1,1]1)\right]
\nonumber \\ & & \times\left(\begin{array}{cc|c}
	1/2   &  1   &  1/2   \\
	S_3   &  i   &  S'_3 
      \end{array}\right) 
\left(\begin{array}{cc|c}
	1/2    &  1   &  1/2   \\
	I_3 &  a   &  I'_3 
      \end{array}\right),
\end{eqnarray}
and
\begin{eqnarray}
\lefteqn{\left\langle [N_c-1,1]1; \frac{1}{2} S'_3; \frac{1}{2} I'_3 \left|  
g^{ia} \right| [N_c-1,1]1; \frac{1}{2}  S_3; 
\frac{1}{2} I_3 \right\rangle =} \nonumber \\ & &
 \left[ \frac{1}{12} \left(K([f']2[f'']2|[N_c-1,1]1)\right)^2-\frac{1}{2} \left(K([f']1[f'']2|[N_c-1,1]1)\right)^2\right] \nonumber \\
& & \times\left(\begin{array}{cc|c}
	1/2    &  1   &  1/2   \\
	S_3    &  i   &  S'_3         
      \end{array}\right)
\left(\begin{array}{cc|c}
	1/2    &  1   &  1/2   \\
	I_3    &  a   &  I'_3 
      \end{array}\right)
\end{eqnarray} respectively. Then Eq. (\ref{sumg}) leads to the following
relation
\begin{eqnarray}
 \lefteqn{\frac{N_c+4}{12}(K([f']2[f'']2|[N_c-1,1]1))^2+\frac{N_c}{6}(K([f']1[f'']2|[N_c-1,1]1))^2 }\nonumber \\ 
& &+\frac{2}{3}\sqrt{\frac{(N_c-3)(N_c-1)}{2}}K([f']2[f'']2|[N_c-1,1]1)K([f']1[f'']2|[N_c-1,1]1)\nonumber \\ 
& & +  \frac{N_c-6}{12}=0,
\end{eqnarray}
which together with the normalization relation (\ref{norm}) forms a system of two 
nonlinear equations for the unknown isoscalar factors. The solution
is exhibited in Table I column 2. The phase convention 
for the isoscalar factors is the same as in Refs. \cite{book,ISOSC}.
 
\subsection{The case S = 3/2, I = 1/2}
These values of $S$ and $I$ imply
\begin{equation}\label{partition2}
[f'] = \left[\frac{N_c+3}{2},\frac{N_c-3}{2}\right], ~~[f''] = \left[\frac{N_c+1}{2},\frac{N_c-1}{2}\right],
\end{equation}
and, as above, there are three nonzero coefficients necessary to express 
the FS state $|[N_c-1,1]1 \rangle$, as linear a combination of S and F
states. In this case, these are $ K([f']2[f'']2|[N_c-1,1]1)$, $K([f']1[f'']2|[N_c-1,1]1)$ and
$ K([f']1[f'']1|[N_c-1,1]1)$. One needs three equations to find the solution.

First, Eq. (\ref{orthog}) gives  the normalization relation
\begin{equation}
(K([f']2[f'']2|[N_c-1,1]1))^2+(K([f']1[f'']2|[N_c-1,1]1))^2
 + (K([f']1[f'']1|[N_c-1,1]1))^2=1.
\end{equation}
Second, Eq. (\ref{sumg}) gives
\begin{eqnarray}
 \lefteqn{\frac{3(3N_c-1)}{10N_c}(K([f']2[f'']2|[N_c-1,1]1))^2+\frac{N_c+1}{2N_c}(K([f']1[f'']2|[N_c-1,1]1))^2}\nonumber \\& & +\frac{3}{N_c}(K([f']1[f'']1|[N_c-1,1]1))^2\nonumber \\
& & -\sqrt{\frac{(N_c-1)(N_c+3)}{5N_c^2}}K([f']2[f'']2|[N_c-1,1]1)K([f']1[f'']2|[N_c-1,1]1)\nonumber \\
& & +\frac{\sqrt{2(N_c-3)(N_c-1)}}{N_c}K([f']1[f'']2|[N_c-1,1]1)K([f']1[f'']1|[N_c-1,1]1)\nonumber \\
& & + \sqrt{\frac{2(N_c-3)(N_c+3)}{5N_c^2}}K([f']2[f'']2|[N_c-1,1]1)K([f']1[f'']1|[N_c-1,1]1) = 1 .
\end{eqnarray}

Third, for the irrep $[f] = [N_c-1,1]$ the Casimir operator identity
becomes (see Eq. (\ref{CASIMIR2}))
\begin{equation}
 \frac{\left \langle S^2 \right\rangle}{2}+\frac{\left\langle T^2 \right\rangle}{2}+2\left\langle G^2 \right\rangle= \frac{N_c(3N_c+4)}{8},
\end{equation}
from which one can derive the following equation
\begin{equation}
 \left \langle s \cdot S_c \right\rangle 
 +\left \langle t \cdot T_c \right\rangle 
 + 4 \left \langle g \cdot G_c \right\rangle = \frac{3N_c-7}{4}.
\end{equation}
Using Eqs (\ref{gGc})--(\ref{tTc}) we obtain
\begin{eqnarray}
 \lefteqn{\frac{3N_c-23}{8}(K([f']2[f'']2|[N_c-1,1]1))^2-\frac{N_c+7}{8}(K([f']1[f'']2|[N_c-1,1]1))^2}\nonumber \\
& & + \frac{\sqrt{5(N_c-1)(N_c+3)}}{4}K([f']2[f'']2|[N_c-1,1]1)K([f']1[f'']2|[N_c-1,1]1) \nonumber \\ 
& & + \frac{1}{2}\sqrt{\frac{5(N_c-3)(N_c+3)}{2}}K([f']2[f'']2|[N_c-1,1]1)K([f']1[f'']1|[N_c-1,1]1) \nonumber \\
& & +\frac{1}{2}\sqrt{\frac{(N_c-1)(N_c-3)}{2}}K([f']1[f'']2|[N_c-1,1]1)K([f']1[f'']1|[N_c-1,1]1)\nonumber \\
& & + \frac{1}{2}(K([f']1[f'']1|[N_c-1,1]1))^2 = \frac{3N_c-7}{4},
\end{eqnarray}
which is the third equation needed to derive the isoscalar factors 
presented in the second column of Table \ref{spin3/2n}. 


\subsection{The case S = 1/2, I = 3/2}
This case is similar to the previous one. The results can be
obtained by interchanging $S$ with $I$.

\section{Fractional parentage coefficients}
We are in the case where the orbital wave function $|[N_c-1,1]pqy \rangle_O $
contains only one excited quark  with 
the structure $s^{N_c-1} p$. 
The one-body fractional parentage coefficients (cfp) help to decouple the 
system into a core of $N_c-1$ quarks and a single quark. 

For $p = 2$ ( obviously $q = 1$ and $y$ is fixed)
the decoupling into a core and a quark gives 
\begin{equation}\label{cfp1}
|[N_c-1,1]2 \rangle_O = \sqrt{\frac{N_c-1}{N_c}} R^c_{[N_c-1]}(s^{N_c-1}) 
R^q_{[1]}(p) - \sqrt{\frac{1}{N_c}}R^c_{[N_c-1]}(s^{N_c-2} p) R^q_{[1]}(s).
\end{equation}
In the first term  the core
is in the ground state, \emph{i.e.} $\ell_c = 0$,  and  is  described by
the symmetric orbital wave function $R^c_{[N_c-1]}(s^{N_c-1})$.
The quark is excited and carries the angular momentum 
$\ell_q = 1$.  In the second term,
the orbital wave function is still symmetric but it contains a unit of 
angular momentum and the quark is unexcited.

For $p = 1$, irrespective of $q$ and $y$ one has
\begin{equation}\label{cfp2}
|[N_c-1,1]1 \rangle_O = R^c_{[N_c-2,1]}(s^{N_c-2} p)  R^q_{[1]}(s)
\end{equation} 
\emph{i.e.} the core is always excited, thus it is described by a mixed symmetric
state,  and the quark is in the ground state.
Denoting  the one-body cfp by $a(p,\ell_c,\ell_q)$ we get
\begin{equation}
a(2,\ell_c=0,\ell_q=1)= \sqrt{\frac{N_c-1}{N_c}},
\end{equation}
\begin{equation}
a(2,\ell_c=1,\ell_q=0)= - \sqrt{\frac{1}{N_c}},
\end{equation}
\begin{equation}
a(1,\ell_c=1,\ell_q=0)= 1.
\end{equation}
to be used in Eq. (\ref{product}).

\section{}
This appendix contains the matrix elements of some operators used to derive 
the isoscalar factors $K([f']p'[f'']p''|[N_c-1,1]1)$ presented in 
Tables \ref{spin1/2n}, \ref{spin3/2n} and \ref{spin1/2delta}.

Let us first recall the general formula of the matrix elements of the generator 
$G^{ia}$ of SU(4) for a given irrep $[f]$. This  is \cite{HP}
\begin{eqnarray}\label{GENsu4}
\lefteqn{\langle [f];I' I'_3;S' S'_3 | G^{ia} |
[f];I I_3;S S_3 \rangle =}\nonumber \\ & & \sqrt{\frac{C^{[f]}(\mathrm{SU(4)})}{2}}   
\left(\begin{array}{cc||c}
	[f]    &  [21^2]   & [f]   \\
	I S   &    11  &  I'S'
      \end{array}\right)_{\rho=1} 
  \left(\begin{array}{cc|c}
	S   &    1   & S'   \\
	S_3  &   S^i_3   & S'_3
  \end{array}\right)
     \left(\begin{array}{cc|c}
	I   &   1   & I'   \\
	I_3 &   I^a_3   & I'_3
   \end{array}\right),
   \end{eqnarray}
where  $C^{[f]}$
is the eigenvalue of the SU(4) Casimir operator for the irrep $[f]$,
the second factor is an isoscalar factor of SU(4),
the third a CG coefficient of SU(2)-spin and the last 
a CG coefficient of SU(2)-isospin.  The isoscalar factors used here are
those of Table A4.2 and A4.5 of Hecht and Pang \cite{HP} adapted to our
notations for $[f]=[N_c]$ and $[f]=[N_c-1,1]$ respectively.
We recall that 
\begin{eqnarray}
C^{[N_c]}(\mathrm{SU(4)}) & = & \frac{3N_c(N_c+4)}{8}, \\
\label{CASIMIR1}
C^{[N_c-1,1]}(\mathrm{SU(4)}) & = & \frac{N_c (3 N_c + 4)}{8}.
\label{CASIMIR2}
\end{eqnarray}
From Eqs. (\ref{fs})--(\ref{isospin}) and (\ref{GENsu4}), one can derive the following matrix elements:
\begin{eqnarray}
 \lefteqn{\langle [N_c-1,1] p;I I'_3;S S'_3 | G_c^{ia} |
[N_c-1,1] p;I I_3;S S_3 \rangle =}\nonumber \\ & & (-1)^{S+I+1}\sqrt{(2S+1)(2I+1)} \sqrt{\frac{C^{[f]}(\mathrm{SU(4)})}{2}}  \left(\begin{array}{cc|c}
	S   &    1   & S   \\
	S_3  &   S^i_3   & S'_3
  \end{array}\right)
     \left(\begin{array}{cc|c}
	I   &   1   & I  \\
	I_3 &   I^a_3   & I'_3
   \end{array}\right) \nonumber \\ 
& & \times \sum_{p',p'',q',q''}(-1)^{S'_c+I'_c}K([f']p'[f'']p''|[N_c-1,1]p)K([f']q'[f'']q''|[N_c-1,1]p) \nonumber \\ 
& &\times \sqrt{(2S'_c+1)(2I'_c+1)}  \left(\begin{array}{cc||c}
	[f]    &  [21^2]   & [f]   \\
	I_c S_c   &    11  &  I'_cS'_c
  \end{array}\right)_{\rho=1}
 \left\{\begin{array}{ccc}
	S   &   1   & S  \\
	S'_c &   1/2   & S_c
   \end{array}\right\}
 \left\{\begin{array}{ccc}
	I   &   1   & I  \\
	I'_c &   1/2   & I_c
   \end{array}\right\},
\label{Gc}
\end{eqnarray}

\begin{eqnarray}
 \lefteqn{\langle [N_c-1,1] p;I I'_3;S S'_3 | g^{ia} |
[N_c-1,1] p;I I_3;S S_3 \rangle =}\nonumber \\ & & (-1)^{S+I+1}\frac{3}{2}\sqrt{(2S+1)(2I+1)}
\left(\begin{array}{cc|c}
	S   &    1   & S   \\
	S_3  &   S^i_3   & S'_3
  \end{array}\right)
     \left(\begin{array}{cc|c}
	I   &   1   & I  \\
	I_3 &   I^a_3   & I'_3
   \end{array}\right) \nonumber \\ 
& & \times \sum_{p',p''} (-1)^{S_c+I_c}
\left(K([f']p'[f'']p''|[N_c-1,1]p)\right)^2
 \left\{\begin{array}{ccc}
	S   &   1   & S  \\
	1/2 &   S_c   & 1/2
   \end{array}\right\}
 \left\{\begin{array}{ccc}
	I   &   1   & I  \\
	1/2 &   I_c  & 1/2
   \end{array}\right\},
\label{g}
\end{eqnarray}

\begin{eqnarray}
 \lefteqn{\langle [N_c-1,1] p;I I'_3;S S'_3 |g\cdot G_c |
[N_c-1,1] p;I I_3;S S_3 \rangle =} \nonumber \\ & & \delta_{S'_3S_3}\delta_{I'_3I_3}(-1)^{S+I+1}\frac{3}{2}\sqrt{\frac{C^{[f]}(\mathrm{SU(4)})}{2}}\nonumber \\
& & \times \sum_{p',p'',q',q''}(-1)^{S_c+I_c} K([f']p'[f'']p''|[N_c-1,1]p)K([f']q'[f'']q''|[N_c-1,1]p) \nonumber \\
& & \times \sqrt{(2S'_c+1)(2I'_c+1)} \left(\begin{array}{cc||c}
	[f]    &  [21^2]   & [f]   \\
	I_c S_c   &    11  &  I'_cS'_c
  \end{array}\right)_{\rho=1}
 \left\{\begin{array}{ccc}
	1/2  &   S'_c   & S  \\
	S_c &   1/2   & 1
   \end{array}\right\}
 \left\{\begin{array}{ccc}
	1/2   &   I'_c   & I  \\
	I_c &   1/2   & 1
   \end{array}\right\},
\label{gGc}
\end{eqnarray}
where $[f]=[N_c-2,1]$ for $p=1$ and $[f]=[N_c-1]$ for $p=2$. $S_c, I_c, S'_c$ and $I'_c$ are determined by
$p', p'', q'$ and $q''$. 

There are also needed
\begin{eqnarray}
\lefteqn{\langle [N_c-1,1] p;I I'_3;S S'_3 | s\cdot S_c |
[N_c-1,1] p;I I_3;S S_3 \rangle =\delta_{S'_3S_3}\delta_{I'_3I_3} (-1)^{S+1/2}\sqrt{\frac{3}{2}}}\nonumber \\
& &\times\sum_{p',p''}(-1)^{S_c} \sqrt{S_c(S_c+1)(2S_c+1)}\left(K([f']p'[f'']p''|[N_c-1,1]p)\right)^2
 \left\{\begin{array}{ccc}
	1/2   &   S_c   & S  \\
	S_c &   1/2   & 1
   \end{array}\right\},
\label{sSc}
\end{eqnarray}
and
\begin{eqnarray}
\lefteqn{\langle [N_c-1,1] p;I I'_3;S S'_3 |t\cdot T_c |
[N_c-1,1] p;I I_3;S S_3 \rangle = \delta_{S'_3S_3}\delta_{I'_3I_3}(-1)^{I+1/2}\sqrt{\frac{3}{2}}}\nonumber \\
& &\times \sum_{p',p''} (-1)^{I_c}\sqrt{I_c(I_c+1)(2I_c+1)}\left(K([f']p'[f'']p''|[N_c-1,1]p)\right)^2
 \left\{\begin{array}{ccc}
	1/2   &   I_c   & I  \\
	I_c &   1/2   & 1
   \end{array}\right\}.
\label{tTc}
\end{eqnarray}





\end{document}